\documentclass[runningheads]{llncs}
\usepackage[dvipdfmx]{graphicx}
\usepackage{sigse}

\begin{document}
\title{Field-sensitive Data Flow Integrity}
\author{So Shizukuishi\inst{1} \and
Yoshitaka Arahori\inst{1} \and
Katsuhiko Gondow\inst{1}}
\authorrunning{S. Shizukuishi et al.}
\institute{Tokyo Institute of Technology}
\maketitle              % typeset the header of the contribution

\begin{abstract}
% サイバー攻撃からソフトウェアを保護する技術は，盛んに研究されている．
% しかし，多くのメモリ保護手法は構造体の要素を正確に保護できない．
% また，構造体の要素を正確に保護するためには複雑なメタデータ管理が必要であり，オーバーヘッドが大きい．
% 本研究は，Field-sensitiveなValue-Flow解析とData-Flow Integrity（DFI）を組み合わせた新しい手法{\OurApproach}を提案する．
% Field-sensitiveなValue-Flow解析により，構造体の要素を区別した解析を行い，正確な解析結果を得る．
% 実行時には，構造体の要素毎に適切にメタデータを管理し，DFIを利用して実行時メモリエラーを正確かつ高速に検知する．
% 提案手法の有効性を検証するため，プロトタイプ実装を用いて既存手法と保護能力や実行時オーバーヘッドを比較する．
% 実験より，既存手法と比較して小オーバーヘッドで正確なデータ保護が可能という結果が得られた．

Although numerous defenses against memory vulnerability exploits have been studied so far, highly-compatible, precise, and efficient defense is still an open problem.
In fact, existing defense methods have at least one of the following problems: they (1) cannot precisely protect structure fields, (2) incur high protection overheads, and/or (3) cannot maintain compatibility with existing code due to imposing memory layout change on the protected program.

In this paper, we propose a novel memory-protection method {\OurApproach} that aims to solve all of these problems simultaneously.
Our key idea is to perform memory protection based on field-sensitive data-flow integrity.
Specifically, our method (1) computes a safe write-read relation for each memory object, at the structure-field granularity, based on field-sensitive value-flow analysis at the compile-time of the protected program.
(2) At run-time, lightweight verification is performed to determine whether each memory read executed by the protected program belong to the safe write-read relation calculated for the memory object at compile time.
(3) This verification is implemented by lightweight metadata management that tracks memory writes at the structure field granularity without changing the memory layout of the target program (especially the structure field layout).
Experiments with our method prototype applied to memory-protection- and performance-benchmarks show that (1) our method is more precise than ASan, a widely-used memory protection tool (in particular, structure protection is more precise), and provides protection that is comparable to Smatus, a state-of-the-art highly-precise memory protection method. 
In addition, our results indicate that the run-time and memory overheads of our proposed method is slightly higher than those of ASan and significantly lower than those of Smatus.
Furthermore, within our experiments, our proposed method was compatible with all of the protected programs.
These results suggest that our proposal is a promising approach to achieve (1) precise, (2) low overhead, and (3) highly-compatible memory protection.
\end{abstract}

% また，プログラムの全データを保護するため，高いオーバーヘッドがかかる．
% さらに，これらのデータ保護を選択的に行い，
% 保護データのみを効率よく保護する．

\section{Background: Control/Data-Oriented Attacks}

While C is a representative programming language for describing systems software including operating systems and network servers, various types of attacks that exploit memory bugs/vulnerabilities in system software in C are still being reported as of 2023.
In most of these attacks, the attacker exploits a memory bug/vulnerability in the victim program to illegally rewrite critical data in memory, thereby intentionally changing the program behavior or causing inconsistencies in the entire data managed by the program.

Attacks that exploit memory bugs/vulnerabilities can be classified into two categories: (1) control-flow-hijacking attacks and (2) non-control data attacks\cite{non-control-data-attack}. 
Control-flow-hijacking attacks take control of the victim program by exploiting its memory vulnerability to illegally rewrite control data such as the target of an indirect-jump instruction, the return address of a function call and the address stored in a function pointer.
Non-control-data attacks illegally rewrite critical data, except control data, such as access permissions and passwords for confidential personal information or system-state variables, thereby causing illegal behavior or unexpected failure of the victim program.

\section{Related Work: Existing Defense Methods}

Numerous protection techniques have been proposed to prevent control-flow hijacking attacks. Among them, Control-Flow Integrity (CFI) has been actively studied as a control-oriented memory protection that is efficient enough to be applied to deployed programs.
CFI monitors control-branch instructions (indirect jumps, function returns and/or indirect calls via function pointers) at run-time to check whether each branch target is a legal instruction-address (control data) or not. The set of legal instruction addresses for each control-branch instruction is typically calculated by performing pointer analysis at compile-time.
Although CFI is an efficient defense enough to be applied to deployed software, it fails to prevent non-control-data attacks because its monitoring is limited to the target address of control-branch instructions (control data).

In order to prevent non-control-data attacks, a large amount of data-oriented memory protection methods have been proposed so far\cite{cyclone,ccured,dfi,asan,effective-san,smatus}. 
Data-oriented memory protection checks, for each memory access at run-time, whether the memory address or the data (includingg non-control one) at the address is legal.
While it incurs higher run-time/memory overheads than CFI, due to increase in the number of monitored data, data-oriented memory protection exhibits higher protection capability, i.e., prevents both non-control-data and control-flow-hijacking attacks.
However, most of existing data-oriented protection techniques are unable to precisely protect structure fields against non-control-data attacks.
A few methods capable of protecting structure fields either perform complex metadata management for structure-field protection, thereby incurring high run-time overhead (for example, Smatus\cite{smatus} a state-of-the-art protection based on pointer metadata), or sacrifice backwards compatibility for lower overheads. 
Compatibility problems arises when, in order to protect structure fields efficiently, the type of pointers and/or structures is changed at compile-time of the target program. 
In other words, as a result of the structure/pointer type change, some parts of existing code fail to work correctly as before.

\section{Problem Summary and Challenges}

At the root of the problems with existing protection methods, are technically difficult challenges:  the design and implementation of data-oriented memory protection that (1) precisely protects structure fields, (2) suppressing run-time overheads, and (3) maintains compatibility with existing code.

\section{Our Approach}

In this paper, we propose field-sensitive DFI, called {\OurApproach}, as a promising approach to solve these technical challenges.
{\OurApproach} combines field-sensitive value-flow analysis and data-flow integrity (DFI)\cite{dfi}, with emphasis on efficient and backward-compatible metadata management at the structure-field granularity. 
{\OurApproach} computes the set of legal definitions (defs) for each use of a structure-field by performing a field-sensitive pointer analysis at compile-time of the target program; similarly, the set of legal defs for each use of a non-structure-field variable.
At run-time of the target program, {\OurApproach} monitors a sequence of defs and uses for each variable (including each structure field).
{\OurApproach} checks, for each use of a variable, if the last-observed definition to be referenced by the use belongs to the set of legal definitions pre-computed at compile time.
If so, it judges that a legal def-use relation holds and continues the execution of the target program; otherwise, it suspends the execution and reports an error, indicating that an illegal def-use relation, i.e., an attack that exploits a memory vulnerability, is observed.
For precise and efficient run-time monitoring of def/use sequence for each variable at the granularity of structure fields efficiently without losing compatibility with existing code, {\OurApproach} adjusts the layout of memory objects allocated by the target program and their corresponding metadata (i.e., last-observed def-location) managed by {\OurApproach}, thereby reducing the amount of metadata required for precise def-use monitoring and speeding up each metadata-lookup without changing the structure type of existing code.

\section{Contributions and Summary of Experimental Results}

We summarize our contributions and preliminary experimental results:
\begin{itemize}
  \item[1.] \textbf{Problem Definition:~} 
    We show that existing defenses against non-control-data attacks suffer from (1) inability to precisely protect structure fields, (2) high run-time overheads, and/or (3) incompatibility with existing code.
  \item[2.] \textbf{Solution:~}
    We propose our idea of combining field-sensitive pointer analysis and DFI to (1) precisely protect structure fields.
    In addition, we propose a novel field-sensitive metadata management, which adjust the layout of both memory objects and their corresponding DFI metadata to (2) reduce runtime- and memory-overheads (3) without changing structure-types.
  \item[3.] \textbf{Experimental Results:~}
    We applied {\OurApproach} to 35 synthetic benchmarks extracted and extended from CBench\cite {finding-cracks-in-shields} and compared our protection capability to existing methods ASan\cite {asan} and Smatus\cite {smatus}. The results show that {\OurApproach} outperforms ASan, especially in protecting structure fields, and that its overall protection capability is promising, although slightly lesser than Smatus.
    Furthermore, the results of overheads evaluation with five benchmarks extracted from SPEC CPU 2006 show that {\OurApproach} imposes +625.8\% runtime overheads on average and a maximum memory overhead of +62.6\%, which are larger than ASan's runtime overhead +59.2 \% and memory overhead +90.9\%.
    However, our overheads are much smaller than Smatus' runtime overhead +1269.8\% and memory overhead +1059.7.9\%.
    In addition, {\OurApproach} was able to maintain compatibility with all target programs.
    These results indicate that {\OurApproach} is an effective approach to provide strong protection at the structure-field granularity with reasonably small runtime/memory overheads, maintaining backward compatibility.
\end{itemize}


\bibliographystyle{splncs04}
\bibliography{references}

\end{document}